\documentclass{basi}

\usepackage{graphicx}
\begin{document}

\title[Globular Clusters]{Some Remarks on Extragalactic  Globular Clusters}

\author[T. Richtler]{Tom Richtler,\thanks{e-mail:tom@mobydick.cfm.udec.cl}\\
$^1$Departamento de F\'{\i}sica, Universidad de Concepci\'on,
Concepci\'on, Chile}
 
\maketitle

\begin{abstract}

I comment (in a review fashion) on a few selected topics  
in the field of extragalactic globular clusters with strong emphasis on recent work.   The topics 
are:
bimodality in the colour distribution of cluster systems, young massive clusters, and the brightest
 old clusters. Globular cluster research, perhaps more than ever, has  lead to important     
(at least to astronomers) progress and problems in galaxy structure and formation.   

\end{abstract}

\section{Introduction}
\begin{figure}[h]
\begin{center}
\includegraphics[width=10cm,angle=-00]{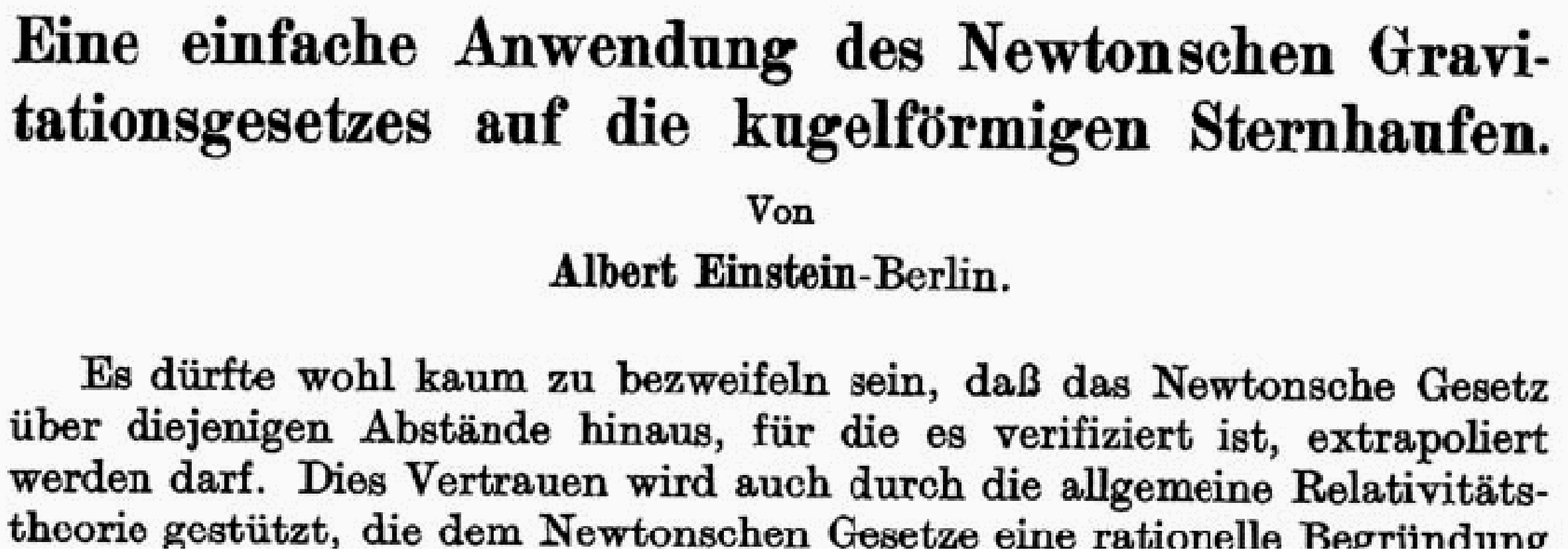}
\end{center}
\caption{The heading of Einstein's paper on M13. The title reads:
A simple application of the Newtonian law of gravitation to globular
star clusters.
}
\label{einstein}
\end{figure}

Since 2005 is the Einstein-year, a talk  on globular clusters can honour it  
by citing Einstein as a pioneer in globular cluster dynamics. In  his paper
on M13 (Einstein 1921) he concluded that the non-luminous mass contributes
no higher order of magnitude to the total mass than does the luminous mass (Fig.~\ref{einstein}). To my
knowledge this has been Einstein's only contact with globular clusters. 
As in other issues, his claim still holds.  \\ 
Globular clusters (GCs) may have the reputation of looking quite similar to
each other, but they are a very inhomogeneous species regarding their
 intrinsic properties. In terms of mass, metallicity, age, density, they span
orders of magnitudes. They are found in all types of galaxies, provided that
the host galaxy's mass is sufficient. Different from what was thought  decades 
ago, GCs are not only survivors from the early Universe, but have also been formed
regularly in galaxies and still form today.\\ 
\indent
The
objective of this contribution is to highlight several recent topics in extragalactic GC research with some
bias towards our own work. It is by no means exhaustive and will bravely
face the usual fate, namely to be outdated soon. A more complete review is
in preparation (Brodie \& Strader 2006).
The field is very extensive, ranging
from stellar populations to cosmology, so only a few points can be illuminated. 
In the following I review/comment on these topics: colour bimodality in globular cluster
systems (GCSs), young massive clusters, and the brightest old clusters.

\section{Colour bimodality}

\begin{figure}
\begin{center}
\includegraphics[width=13cm,angle=-00]{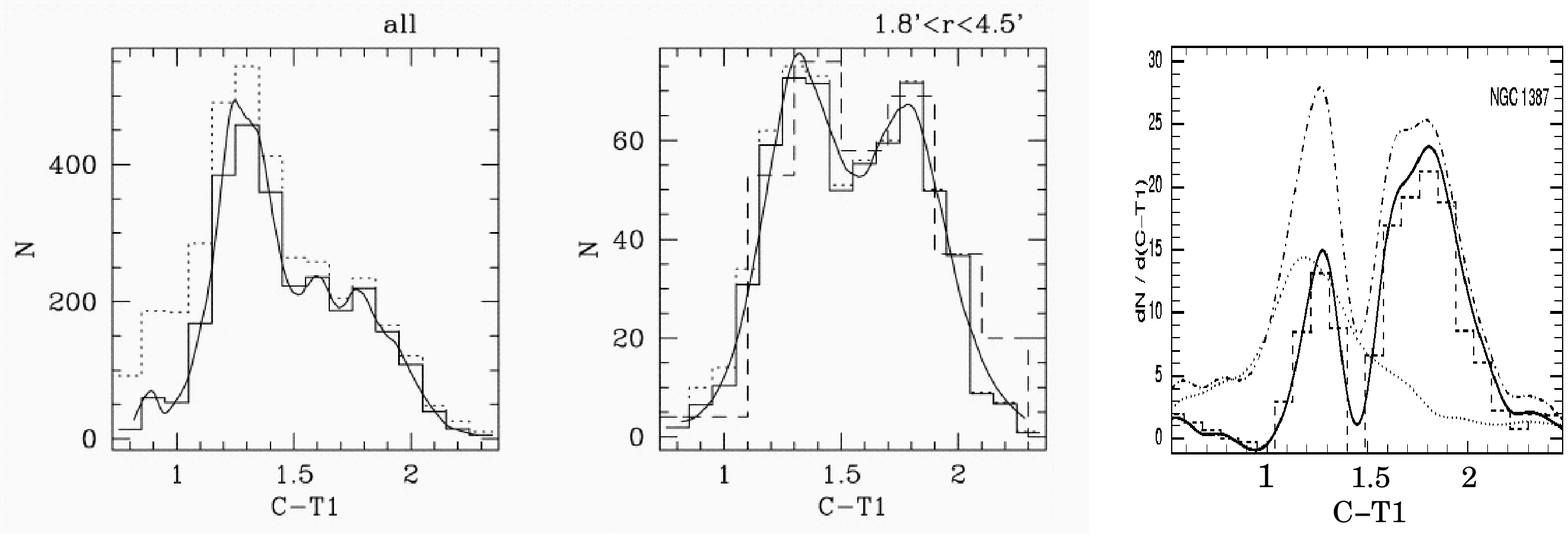}
\end{center}
\caption{The colour distribution of the GCSs of NGC 1399 and NGC 1387 in the
Washington system (Dirsch et al. 2003, Bassino et al. 2005). The left panel
shows the entire system of NGC 1399, the middle panel its inner part. 
The solid line in the left panel is a smoothing of the solid histogram which
displays the background corrected counts. The dotted line indicates the total
counts. The same holds for the middle panel, except that the dashed line shows the counts of Ostrov et al. 
(1998). 
The right panel shows the GCS of NGC 1387, where the solid line is a smoothing 
of the dashed
histogram, giving the background corrected counts. The dashed-dotted line means
the smoothed total counts and the dotted line the background, which 
is
dominated by the GCS of NGC 1399.
The colour distributions are clearly bimodal, and that of NGC 1387 unique in its
very large ratio of red to blue globular clusters.
}
\label{bimod}
\end{figure}

\begin{figure}
\begin{center}
\includegraphics[width=8cm,angle=-00]{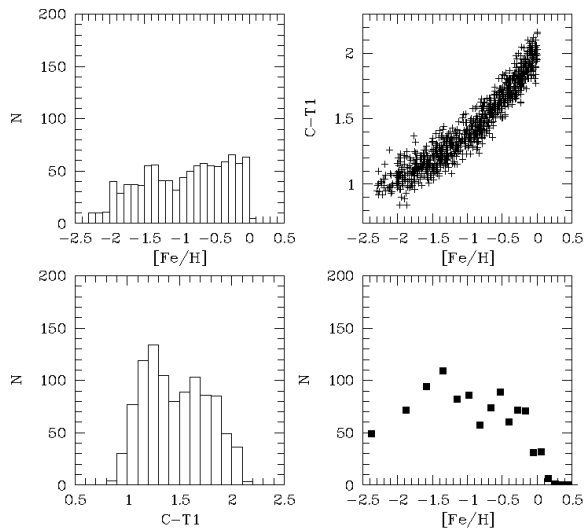}
\end{center}
\caption{ This plot illustrates for the Washington system that a non-peaked metallicity distribution can
result in a bimodal colour distribution. The upper left panel is the original
metallicity distribution, the upper right panel the adopted metallicity-colour relation
 (Harris \& Harris 2002) with a Gaussian scatter of 0.08 mag around the
mean relation to simulate
second parameters and photometric errors. The lower left panel shows the colour histogram, now bimodal, and the lower right panel the corresponding metallicity distribution.
}
\label{simulation}
\end{figure}

There is no doubt that the GCSs of early-type galaxies
exhibit substructure. The distinction between  halo and bulge clusters with their
differences in spatial distribution, chemical composition and kinematics is already familiar
from the Galactic system. However, in the case of early-type galaxies, it has
been the colour distribution which first indicated subtructure. Ashman \&
Zepf (1992) were the first to investigate this. Because they 
identified two peaks in the colour distributions of the GCSs of NGC 5128 and
NGC 4472 (Zepf \& Ashman 1993), they termed them ''bimodal'' (in fact, they plot metallicities, not
colours). A common scenario for the formation
of giant ellipticals is through the merging of smaller galaxies. Linking
bimodality with the merger scenario, they predicted the abundant formation
of GCs in mergers, following Schweizer (1987). These newly formed GCs, presumably enriched, constitute the
red (metal-rich) peak of the bimodal distribution. Later on, it has been found
that indeed a lot of new GCs are formed in the star bursts which accompany 
merger events (see references in section 3). Bimodality could thus be understood as the sign of two epochs
of GC formation. Other models with different prescriptions followed. Forbes 
et al. (1997) explained bimodality by having the metal-poor clusters
formed ''in situ''. C\^ot\`e et al. (2001) found the colour distributions
most consistent with the accretion/infall of metal-poor clusters and the formation
of metal-rich clusters in a dissipational collapse.
Beasley et al. (2002) considered cluster formation within a semi-analytical model of galaxy
formation and obtained bimodal metallicity distributions, the metal-poor clusters   
stemming from protogalactic fragments, and the metal-rich ones from subsequent
gaseous mergers.
More  observations in the Washington system and
in V-I (mainly based on HST data, Kundu \& Whitmore 2001a, Kundu \& Whitmore 2001b, Larsen et al. 2001) demonstrated that the majority of early-type galaxies, also including fainter galaxies, show 
bimodal colour distributions. In fact, {\it all} GCSs observed so far in the Washington 
system revealed bimodal distributions, while in V-I, perhaps due to the inferior
metallicity resolution, a certain fraction remained unimodal or unclear.
Interesting cases are NGC 524 or NGC 4365 (Larsen et al. 2001)
which have quite rich GCSs, but appear unimodal. It would be interesting to observe these
GCSs also in the Washington bands in order to see whether this is simply an effect of the photometric
system.\\
\indent
However, one should note that in many cases, the term ''bimodal'' does not result from a clear-cut
visual appearance in the colour distribution, but from a statistical analysis,
often done with a KMM test.\\
\indent
Figure~\ref{bimod} shows the colour distributions in the Washington system of the GCSs of
NGC 1399, the central giant elliptical in the Fornax cluster, and NGC 1387, a Fornax S0 galaxy near  NGC 1399 (Dirsch et al. 2003, Bassino et al. 2005). To my knowledge
this is the clearest bimodality seen in any galaxy. Being of quite low
luminosity and an S0-galaxy, NGC 1387 is not a good candidate for supporting
the merger scenario. However, the processes which produce S0s have, according
to common wisdom to do, with interactions of galaxies, and the predominance
of red clusters might have its explanation in the star burst which took place
while NGC 1387 was transformed into an S0 galaxy.   
If also low-luminosity ellipticals have bimodal colour distributions, as seen e.g. in
NGC 1427 by Forte et al. (2001) one may
guess that all(!) GCSs have that property, as long as the full colour range is sufficiently 
sampled.\\ 
\indent
Is there a visible systematic relation between the peak locations and 
the properties of the host galaxy? Larsen et al. (2001), Kundu \& Whitmore (2001a), and
Kundu \& Whitmore  (2001b) agree that the red peak is redder for host galaxies of higher luminosity.
This is not so clear for the blue peak. 
Larsen et al. (2001) also find  the blue peak 
becoming bluer with decreasing galaxy luminosity and/or central velocity dispersion. 
Kundu \& Whitmore, using partly the same data set, failed in uncovering such a 
relation. I shall come back to this shortly.
\indent
In the Washington system, however, where the bimodalities show up much better
than in V-I, such a correlation for the blue peak is less obvious (Bassino et al. 2005). This might have its basis in the
fact that the HST data are much more homogeneous than ground based 
photometries from different telescopes, authors, reduction methods etc.. However, it also can mean that the
colour of the blue peak is in fact (more or less) universal (of course, it should not be strictly universal:
if the host luminosity is so low that the cluster system consists only of a handful of metal-poor clusters,
it is plausible that the mean colour becomes bluer at some point).\\
\indent
An effect which argues for an almost universal blue peak and which has not yet been discussed before in the
context of bimodalities is illustrated in Fig.~\ref{simulation}. In the upper left
panel a non-peaked distribution of cluster metallicities is shown (drawn from an exponential distribution). To transform
it into colours we apply the relation of Harris \& Harris (2002).
However, at a given metallicity, the colour can vary due to second parameter
effects and for real data also due to photometric errors. We simulate the combined
effect by a Gaussian scatter with a dispersion of 0.08 mag. One also notes
the non-linearity in the upper right panel. Then we rebin in colour (lower left
panel). The striking blue peak is a consequence of the scatter
around a mean colour-metallicity relation in combination with the declining metallicity
resolution at bluer colours. The lower right panel shows the rebinning in metallicity
now assuming a unique metallicity-colour relation. The result differs significantly
from the input distribution. \\
\indent
An intrinsically bimodal metallicity distribution plausibly gives a bimodal colour distribution, but our example shows that also a unimodal metallicity distribution may
result in a bimodal colour distribution. The colour of the blue peak is
uniform, depending on the (uniform) colour-metallicity relation and the scatter
 around the mean relation. Any non-linearity would help to create a blue peak.
Strader et al. (2004) find that the mean colour of metal-poor clusters (they quote [Fe/H] $<$ -1) follows an extremely shallow relation
like $\mathrm (V-I)_{mean} \sim -0.01 \cdot M_V$ ($\mathrm M_V$ being the host galaxy luminosity). They
include intrinsically faint galaxies hosting only a few clusters, where the metallicity distribution is not
fully sampled. Therefore their finding does not necessarily contradict a universal blue peak in the above
sense. But Strader et al. (2005a) and Peng et al. (2005) found a similar slope among early-type galaxies in Virgo, so
the blue peak does not seem to be strictly universal. 
Whether the described effect dominates the colour distribution is left to a deeper
analysis, but at least it could be responsible for the shallowness of the above relation. Metallicity distribution and colour distribution might be
less closely linked than previously thought. However, a colour distribution like that seen for NGC 1387 
cannot be the result of non-linearity or second parameters, but probably indicates the disk and halo structure of an S0-galaxy, as has been found also for NGC 1380 (Kissler-Patig et al. 1997).
\indent
Within a merger scenario, the resulting GCS is expected
to be a composit of many different cluster formation processes: GC formation in the star bursts of the
merger components, in tidal  tails, the original clusters of the merger components, and late infall of clusters. 
The majority of clusters in early-type galaxies must have formed at high redshift: Strader et al. (2005b) found in their sample of spectroscopically derived 
ages and metallicities no clusters, whether blue or
red, which are younger than about 10 Gyr. Moreover, no age difference between
metal-poor and metal-rich clusters is visible within the uncertainties. The mere morphological phenomenon of colour bimodality
as an indication of two and only two formation epochs should thus not be overinterpreted. 
Cluster colours provide  first hints, but for the physically interesting substructure of a GCS to be revealed,
one needs spectroscopic metallicities, ages, and kinematics.\\ 
\indent
When speaking about colour bimodality, one should also mention where it does
{\it not} occur, namely among the brightest clusters in rich cluster systems
(Dirsch et al. 2003, Harris et al. 2005). The absence of metal-poor objects
among very massive clusters might be explained by massive progenitor clouds
which are systematically more enriched than their less massive counterparts
(Harris et al. 2005). Another (not necessarily competing) possibility is that this mass range is contaminated by tidally stripped dwarf galaxies whose nuclei
now appear as globular clusters (compare also section 4).

\section{Young massive clusters}

The classic view on GCs as representatives of the oldest stellar populations
has experienced considerable modifications. Thirty years ago it was an
irritating fact that the Magellanic Clouds host GC-like objects which are,
however, much younger than the old Galactic GCs (e.g van den Bergh 
1975). Nowadays we know that the conditions for the formation
of GCs were not exclusively realized in the early Universe but that GC formation is a regular feature of star formation processes in a variety of environments,
the most spectacular ones being galaxy mergers (e.g. Whitmore et al. 1993, Whitmore \& Schweizer 1995, de Grijs et al.
2003, Tran et al. 2003).
The most massive young cluster known today is W3 in the merger galaxy NGC 7252
with an age of about 500 Myr and a mass of about $\rm 8 \cdot 10^7 M_{\odot}$ (Maraston et al. 2004). 
It has an absolute magnitude of $\mathrm M_V = -16.3$ and thus rivals M32, the compact companion of the Andromeda galaxy. It still will have about $\mathrm M_V = -13$ when it is 12 Gyrs old, and thus may be a progenitor for ''Ultracompact
Dwarfs'' (see the next section). Clusters often tend to form in complexes which for example are found in the Antenna galaxy (Whitmore et al. 1999).
The simulations of Fellhauer \& Kroupa 
(2002)
predict a rapid merging of star cluster 
complexes, so consequently 
 Fellhauer \& Kroupa (2005) propose the merging of clusters as a possible formation scenario for W3.\\
\indent 
 Apparently, merger events
provide the best conditions for forming massive clusters. 
If this is so one would expect to find intermediate-age globular clusters
in galaxies which are known to be merger remnants. Nearby examples, where
indeed intermediate-age clusters have been found, are
Centaurus A,  NGC 1316, and NGC 3921 (Peng et al. 2004, Goudfrooij et al. 2004, Schweizer et
al. 2004). Intermediate-age clusters have also been detected photometrically in early-type galaxies
which are not obvious merger remnants (Puzia et al. 2002, Hempel \& Kissler-Patig 2004). 
The number of clusters
formed in a merger event may plausibly depend much on its specific properties
 including the possibility that no clusters or only an insignificant number are 
formed. NGC 1052 shows some signatures of a recent merger with an associated star burst, but
 Pierce et al. (2005) 
did not find evidence for younger ages  in their sample of 16 GCs.\\
\indent 
Also starburst galaxies possess populations of young massive clusters. 
One example is NGC 1569 (Hunter et al. 2000) but there are many more.\\
\indent 
However, the formation of globular clusters is not restricted to these relatively
extreme environments. Larsen \& Richtler (1999, 2000)
 performed the first systematic search among spiral galaxies for young luminous
clusters. In their sample of 21 nearby face-on spirals, they found many. 
Introducing the parameter $\mathrm T_L$, which measures the total light coming from
the identified young clusters normalized to the light of the host galaxy, they
showed that this parameter correlates best with the far-infrared luminosity of the
host galaxy. This in turn is a measure for the star formation rate.\\
\indent 
In comparison with the sample galaxies, the Milky Way's star formation rate is quite low, and
the fact that Galactic massive young clusters are rare or even not existing, is therefore not
suprising. 
However, one needs dynamical information in order to assess whether the luminous clusters found
 in galactic disks really are the young 
counterparts of old GCs. Larsen \& Richtler (2004) estimated virial masses for two bright clusters in
M83, resulting in $\mathrm 4.2 \cdot 10^5 M_{\odot}$ and $\mathrm 5.2 \cdot 10^5 M_{\odot}$, respectively.
The cluster ages are about $\mathrm 10^8$ and $\mathrm 10^7$ years. The derived  masses are consistent
 with ''normal'' stellar mass functions and show that these objects deserve
the label ''young globular clusters''.\\
\indent   
\begin{figure}
\begin{center}
\includegraphics[width=3cm,angle=-00]{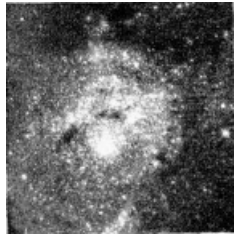}
\end{center}
\caption{The image shows a ''peculiar'' complex in NGC 6946 (Larsen et al. 2002).
A young massive cluster is located near the center of a region of enhanced star formation
of almost circular shape. The diameter is about 600 pc. A number of smaller star clusters
are found as well.
}
\label{N6946}
\end{figure}
In the sample of Larsen \& Richtler, NGC 6946 is worthy of particular note.
 Figure~\ref{N6946} shows an interesting
cluster and its environment in this galaxy, investigated in detail by Elmegreen et al. (2000), Larsen et al. (2001), Larsen et al. (2002),
 and Efremov et al. (2002).  The cluster is located near the center of a stellar complex with very
 high surface brightness and an approximately
circular shape (diameter 600 pc) at a distance of 4.8 kpc from the center of  NGC 6946. Its brightness
 is $\mathrm M_V = -13.2$
and its mass is estimated to be about $\mathrm 5 \cdot 10^5 M_\odot$. Other fainter clusters could be
identified in this complex. Moreover, dust features are visible. It may be spherical, but a face-on disky structure with the massive cluster
as the central object is plausible, regarding the thin disks even in large spiral galaxies.
Although by far not as massive as W3, this concentrated occurrence of star clusters in a region with a high
star formation rate is perhaps an example for the scenario suggested by Fellhauer \& Kroupa
(2002), leaving a cluster embedded in an extended envelope.\\
\indent
What does the correlation of star formation rate and efficient clustering mean? Could a high star formation
rate be a trigger for cluster formation or do both result from the same physical conditions? It
seems that the latter is the case. Cluster formation needs regions of high pressure (Elmegreen \& Efremov 1997), 
the natural
sites being the cores of giant molecular clouds. A high density of cold gas results in a high star formation
efficiency and favours clustering (Geyer \& Burkert 2001). In these regions the star formation rate is also
 high. Kravtsov \& Gnedin (2005) simulate cluster formation in the context of hierarchical cosmologies and find 
a similar relation between the efficiency of forming massive clusters and the star formation rate as found by
Larsen \& Richtler. In their model, the interpretation simply is the above.  
However, the conditions which lead to a large reservoir of cold, dense gas might be different in different
galaxy types. 
 Billett et al. (2002) searched for  massive clusters in dwarf irregular galaxies. The main result is that dwarf galaxies
do not follow the trends observed for spirals.  Billett et al. explain the correlation between star formation rate and clustering found for
large spirals by correlations between star formation rate, maximum cluster mass, column density, and pressure in
 combination with statistical effects. They suggest that GC formation
in dwarf galaxies occurs under different circumstances than in spiral galaxies, triggered by large scale
flows in the absence of shear. The most massive clusters in dwarf galaxies are statistically not expected, but probably
need unique conditions. 
It therefore seems that  a general high star formation rate is a necessary condition for GC formation, but not a sufficient one.

\section{The brightest old clusters}

\begin{figure}
\begin{center}
\includegraphics[width=14cm,angle=-00]{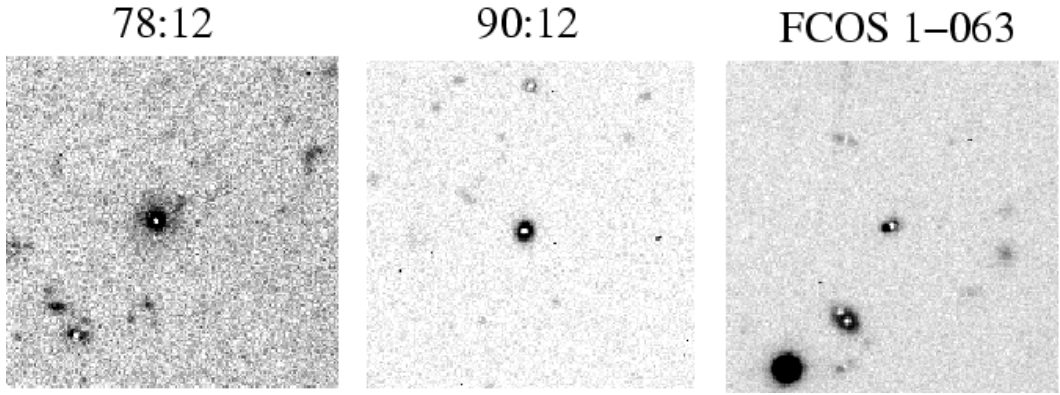}
\end{center}
\caption{PSF-subtracted images for some resolved objects near NGC 1399 (Richtler et al. 2005).
}
\label{nainital}
\end{figure}

The long history of research on the brightest cluster in the Galactic system, $\omega$ Centauri,
has yet not converged towards a commonly accepted picture. The last two years saw some surprising
discoveries (see Piotto et al. 2005 and references therein). One could cautiously say that a scenario in which  $\omega$ Centauri has 
formerly been the nucleus of a dwarf galaxies which has fallen into the Milky Way and has been
tidally stripped, is perhaps the most promising working hypothesis (e.g. Hilker \& Richtler 2000).
 Other models are, however,
not ruled out. For example, Fellhauer \& Kroupa (2003) discuss the formation 
of  super-star clusters in an ancient star burst, which subsequently merged to form 
 $\omega$ Centauri. \\
\indent 
If it is already that difficult to understand a Galactic GC with all possibilities of detailed
investigation, it is not surprising that unfamiliar objects in external galaxies are even more
difficult to understand. 
Hilker et al.
(1999), searching for new members of the Fornax galaxy cluster, ''discovered'' two GC-like 
objects with radial velocities placing them into the Fornax cluster, but with magnitudes brighter than any GC known until then. In fact, the brighter one was resolved and labelled ''compact dwarf elliptical''
earlier by Minniti et al. (1998).  They already 
noted that these objects could be the stripped nuclei of dwarf galaxies, as was suggested before
by the simulations of Bassino et al. (1994). Drinkwater et al. (2000) and Phillips et al. 
(2001) found three more in the course of the 2dF survey and introduced the designation
"Ultra Compact Dwarfs (UCDs)". These objects have absolute magnitudes
in the range $\mathrm -12.1 < M_V < -13.4$. HST observations and VLT spectroscopy permitted to derive
structural parameters and M/L-ratios (Drinkwater 2003). Their half-light radii are
in the range 15-30 pc, much larger than those of most GCs. $\mathrm M/L_V$-values lie between 2 and 4,
not strikingly high for GCs. Metallicities are not well known.
 For one object, UCD 2, spectral line indices are available (Mieske et al. 2004)
and their metallicity of -0.6 dex agrees well with that derived from its Washington colour (Richtler
et al. 2005), when an old age is assumed.\\
\indent   
All these objects have been found within 30 arcmin from NGC 1399. This is still within the area covered by the
GCS of NGC 1399, which extends to over 40 arcmin (Bassino et al. 2006). No object
of this
kind  has been found in the general Fornax field, which already suggests that the close distance
to NGC 1399 plays an important role.\\ 
\indent
\begin{figure}[ht]
\begin{center}
\includegraphics[width=13cm,angle=-90]{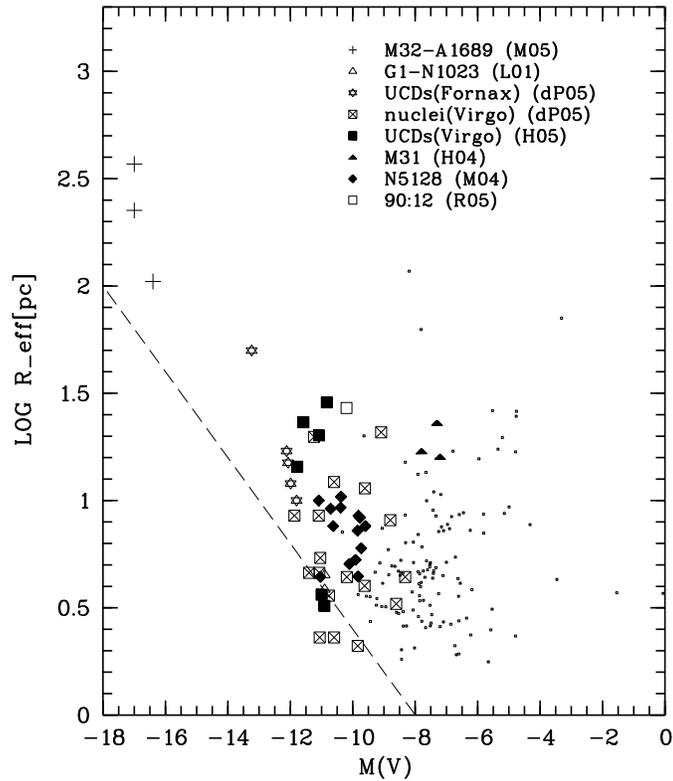}
\end{center}
\caption{Absolute visual magnitudes vs. effective radii (pc) for a number
of objects. The dashed line is a line of constant surface brightness within the
effective radius ($\mathrm m_V = 14.8/arcsec^2$). References: dots (Galactic GCs): Harris 1996; M05: Mieske et al. 2005; L01: Larsen 2001;
dP05: de Propris et al. 2005; H05: Ha\c{s}egan et al. 2005; 
H04: Huxor et al. 2004; M04: Martini \& Ho 2004; R05: Richtler et al. 2005.
}
\label{compacts}
\end{figure}
The UCDs populate a previously deserted area in the $\mathrm M_V - r_{eff}$-plane. Meanwhile, more
objects with similar characteristics have been found in the Fornax cluster, but also in other
environments. Unfortunately, HST-images for Fornax objects exist only for the 7 UCDs, so structural
parameters remain uncertain, although it is possible to resolve larger GC clusters also on ground
based images in good seeing. Richtler et al. (2005) present a short list of "noteworthy"
objects which have been found in the course of a spectroscopic survey of the NGC 1399 GCS (Richtler
et al. 2004, Dirsch et al. 2004).  Figure~\ref{nainital} shows a selection of such objects. On VLT images, some of them are well
resolved and one (90:12) has an effective radius of 27 pc, rivaling  those of UCDs, but an absolute
magnitude of ''only'' $\mathrm M_V \approx -10.3$. Another one (78:12) has a large, faint halo with a radius
of at least 200 pc, which will probably even be larger on deeper images. Such an object has never been seen before,
 although ''extra-tidal'' light has been found as well around GCs in NGC 5128 (Harris et al. 2002).
Since all these objects have been found on only two FORS2 fields (in total 42 $\mathrm{arcmin^2}$), one expects many more to be discovered
in a complete census of the system.
Mieske et al. (2004) identify more bright GC-like objects, which still await HST- or ground based
high resolution imaging.
Meanwhile, a lot of interesting objects have been found in other galaxies as well. Since there
is no space for a thorough discussion, I mention briefly the most recent ones.
Huxor et al. (2004) found extraordinarily extended clusters in the halo of M31 with absolute
magnitudes around $\mathrm M_V \approx -7.2$ and half-light radii of about 30 pc. Mieske et al.
(2005) discovered twins to M32 in the galaxy cluster Abell 1689, leaving M32 no longer
lonesome in its properties.  
Ha\c{s}egan et al. (2005), in the course of the ACS Virgo survey, discovered bright GC-like
objects around M87 and other Virgo galaxies, but with half-light radii resembling those of the
UCDs in the Fornax cluster. Martini \& Ho (2004) present a list of 14 bright clusters in
NGC 5128 with measured velocity dispersions and structural parameters (see also G\'omez et al. 2005). The
reader is also referred to the recent results of the ACS Virgo survey (Jordan
et al. 2005), which we cannot discuss here.\\
\indent
Is any combination of luminosity and concentration possible? Figure~\ref{compacts} plots (following
Huxor et al. (2004) absolute magnitudes
and effective radii for GCs and GC-like objects, including Galactic GCs, dwarf galaxy
nuclei, and M32-like objects.
Identifications and references are given in the figure and figure caption. Population properties
like ages and metallicities are not known for most of the newly discovered objects, so normalisation
to, say, solar abundance and an old age is not possible. 
However, it seems that the indicated dashed
line (line of constant surface brightness within an effective radius) constitutes some sort of
 limit to the surface brightness, corresponding to $\mathrm \mu_V = 14.8/arcsec^2$. Until now, the brightest UCD in Fornax (UCD 3)
seems to be unique; the others finding their counterparts also around M87 in the Virgo cluster. While
UCD 3 seems to be particular, the others might well be understood as the brightest GCs
in the NGC 1399 system. In Fig.~\ref{compacts} there  is only a small morphological gap (if any) 
between UCDs and ''real'' GCs (whatever they are) and one may expect  that the space towards
fainter magnitudes, but equally large effective radii, will be filled once a complete census of
NGC 1399 GCs will be available.  
Bekki et al. (2001, 2003)  simulated the stripping of nucleated dwarf galaxies 
in the potential of a giant galaxy (''galaxy threshing''). These simulations support the scenario
in which UCDs emerge from nucleated  dwarfs. However, the nuclei themselves are not greatly affected
in the simulations so one should expect to find nuclei with similar properties as the UCDs. 
De Propris et al. (2005) compare structural properties of the nuclei of Virgo dwarf 
 galaxies with those of UCDs and conclude that most dwarf nuclei are too faint and too concentrated
to be progenitors for UCDs. As can be seen in Fig.~\ref{compacts}, only two or three nuclei could be candidates
for fainter UCDs. The brightest one, UCD 3, shows a surface brightness profile different from the others,
which are well fit by King profiles (Drinkwater et al. 2003). An additional exponential halo
with a scale length of 60 pc is required to obtain a good fit. Adopting an old age, UCD3 must have been
about 4 mag brighter at the time of its formation. This combination of brightness and concentration is 
extremely rare, but a plausible intermediate-age counterpart is W3 in NGC 7252 (Maraston et al. 2004). 
\\ 
\indent
There are more arguments in favour and against dwarf galaxies as progenitors for UCDs.
 An interesting idea is from Mieske et al. (2004) who expect
a dynamical difference between ''normal'' GCs and objects which are the result of ''galaxy threshing''.
 Since the stripping process relies on elongated orbits and small perigalactic radii (Bassino 1994, Bekki et al. 2001) the velocity dispersion of UCDs should be lower 
 than that of GCs because one selects a radially biased subpopulation and one expects to find UCDs
far from their pericenters. The number of UCDs
is admittedly small for dynamical analyses. However, from the sample of Mieske et al. it appears that  
the brightest GCs have indeed a (marginally) smaller velocity dispersion.\\
\indent
Perhaps a deeper insight into the nature of UCDs results from the work of Ha\c{s}egan et al. (2005). They present a list of bright and relatively compact objects around M87 and other galaxies in
the Virgo cluster, which they name ''Dwarf-Globular Transition Objects (DGTOs)''. For six of them,
internal velocity dispersions from Keck spectroscopy are available. Two of them
are extremely compact and have structural properties similar to G1 in M31 or Larsen's object in
NGC 1023 (Larsen 2001). Their $\mathrm M/L_V$-ratios are high (about 3) but still consistent with old, metal-rich
GCs. Three of the other four, however, have $\mathrm 6 < M/L_V < 9.4$. Interestingly, the object
with the largest effective radius again has $\mathrm M/L_V \approx 3$. Ha\c{s}egan et al. propose that
the high $\mathrm M/L_V$-ratios are due to a high dark matter content which is the distinguishing
property between UCDs and ''normal'' GCs. The dark matter halo would then be the dark matter debris
of a stripped dwarf galaxy. If this was true, none of the Fornax objects would be
real UCDs in the sense of Ha\c{s}egan et al., since the M/L-values measured by Drinkwater
 (2003) are all plausibly consistent
 with stellar populations.\\ 
\indent
However, since the measured velocity dispersion essentially is the central velocity dispersion, the mass profile of UCDs with a dark matter halo must be closely following the light profile to have a dark matter dominated
central region. This again would be strange, because stripping of dwarf galaxies works most effectively with a 
flat dark matter core (Bekki et al. 2003). Another possibility is a central black hole,
which would have to be rather massive to increase the M/L-ratio by a factor of 2. Baumgardt et al.
(2005)
 show that a black hole can puff up the core of its hosting cluster quite
considerably.
 Einstein's finding for M13 (see the introduction) perhaps is still
 relevant for  
distinguishing globular clusters from UCDs.\\
\indent
It is also interesting to note that many of the brightest globular clusters, including W3, have quite high
 ellipticities (e.g. Larsen 2001), 
while the UCDs and dwarf galaxy nuclei appear more spherical. If the ellipticity is due to rotation (which is not known), 
one could think of it as a consequence of angular momentum conservation in the merging of smaller
subunits, but the simulations of Fellhauer \& Kroupa (2002) reveal quite spherical
objects as a result of cluster merging. \\
\indent 
However, the example of W3 shows that very massive clusters can form without invoking a scenario with infalling 
donator galaxies and it remains open to what degree infalling dwarf galaxies can shape the appearance of
a GC system.  

\section{Concluding remarks}
  
The field of GC research, of which a few topics have been presented here, is as exciting as ever.
 One cannot yet claim that the formation of globular clusters
and their relation to their host galaxies are well understood,  neither does it 
constitute a total mystery. Globular clusters may even have formed in different ways as
entire objects or by coagulation of smaller units.
Their potential for understanding galaxy formation and evolution is just on the verge of being
exploited.
The determination of ages, metallicites, internal dynamics, and kinematic and dynamical properties 
of cluster systems requires lots of work. However, it is fun by itself and promises deep insights into the
evolution of structure. Einstein's notion that M13 does not host large amounts of dark matter 
touched already upon one of the fundamentals of structure formation and today even appears more revelant
than 85 years ago. 
  
\section{Acknowledgements}
 I thank the Aryabhatta Research Institute of Observational
Science, Naini Tal, for the invitation, warm hospitality (which more than balanced the low
temperatures), and financial support. Financial support  
from the Chilean {\sl Centro de Astrof\'\i sica} FONDAP No. 15010003
is also gratefully acknowledged.  
I am indebted to Boris Dirsch for many discussions during the last years. That
bimodalities can also emerge from non-peaked metallicity distributions, was originally
his idea. I thank Ylva Schuberth, Doug Geisler, Michel Hilker, S{\o}ren Larsen and an
anonymous referee
 for critical reading and comments.


\newpage
\begin{thebibliography}{10}
\bibitem {} Ashman, K., Zepf, S. 1992, ApJ, 384, 50 
\bibitem {} Bassino, L.P., Muzzio, J.C., Rabolli, M. 1994,
ApJ, 431, 634
\bibitem {} Bassino, L.P., Richtler, T., Dirsch, B. 2005, MNRAS, in press (astro-ph/0511770)
\bibitem {} Bassino, L.P., Faifer, J.C., Forte, J.C. et al. 2006, A\&A, in press

\bibitem {} Baumgardt, H., Makino, J., Hut, P. 2003, ApJ 589, L.25
\bibitem{} Baumgardt, H., Makino, J., Hut, P. 2005, ApJ 620,
238
\bibitem{} Beasley, M. A., Baugh, C.M., Forbes, D.A. et al. 2002, MNRAS, 333, 383

\bibitem{} Bekki, K., Couch, W.J., Drinkwater, M. J. 
2001, ApJ, 552, L105

\bibitem{} Bekki, K., Couch, W.J., Drinkwater, M. J., Shioya, Y.
2003, MNRAS, 344, 399
\bibitem{} Billett, O.H, Hunter, D.A., Elmegreen, B.G. 2002, AJ, 123, 1454

\bibitem{} Brodie, J. P., Strader, J. 2006, ARAA, in preparation

\bibitem{} C\^ot\`e, P, McLaughlin, D., Hanes, D. et al. 2001,
ApJ 559, 828

\bibitem{} de Grijs, R., Lee, J.T., Clemencia Mora Herrera, M. et al. 2003, New A. 8, 155

\bibitem{} De Propris, R., Phillips, S., Drinkwater, M.J. et al. 2005,
ApJ, 623, L105

\bibitem{} Dirsch, B., Richtler, T., Geisler, D. et al. 2003, AJ 125, 1908

\bibitem{} Dirsch, B., Richtler, T., Geisler, D., et al. 2004, AJ 127, 2114

\bibitem{} Drinkwater, M., Jones, J.B., Gregg, M.D., Phillipps, S. 2000, PASA, 17, 227

\bibitem{} Drinkwater, M., Gregg, M.D., Hilker M., et al. 2003, Nature, 423, 519 (see also astro-ph/0306026)

\bibitem{} Einstein, A. 1921, Festschrift der Kaiser-Wilhelm Gesellschaft
zur F\"oderung der Wissenschaften zu ihrem zehnj\"ahrigen Jubil\"aum dargebracht von
ihren Instituten, Springer, Berlin 1921, p. 50

\bibitem{} Efremov, S.U., Pustilnik, S.A., Kniazev, A.Y. et al. 2002, A\&A, 389, 855

\bibitem{} Elmegreen, B. G., Efremov, Y. 1997, ApJ, 480, 235

\bibitem{} Elmegreen, B. G., Efremov, Y., Larsen, S.S. 2000, ApJ, 535, 748


\bibitem{} Fellhauer, M., Kroupa, P. 2002, MNRAS, 330, 642

\bibitem{} Fellhauer, M., Kroupa, P. 2003, Ap\&SS, 284, 643

\bibitem{} Fellhauer, M., Kroupa, P. 2005, MNRAS, 359, 223

\bibitem{} Forte, J.C., Geisler, D., Ostrov, P. et al. 2001, AJ, 121, 1992

\bibitem{} Forbes, D.A., Brodie, J.P., Grillmair, C.J. 1997, AJ 113, 1652

\bibitem{} Gebhardt, K., Rich, R., Ho, L.C. 2002, ApJ, 578, L41

\bibitem{}  Geyer, M.P., Burkert, A. 2001, MNRAS, 323, 988

\bibitem{} G\'omez, M., Geisler, D., Harris, W.E. et al. 2005, A\&A, in press
(astro-ph/0510544)

\bibitem{} Goudfrooij, P., Gilmore, D., Whitmore, B.C., Schweizer, F. 2004, ApJ, 613, L121

\bibitem{} Harris, W.E. 1996, AJ, 112, 1487

\bibitem{} Harris, W.E., Harris, G.L.H., 2002, AJ, 123, 3108

\bibitem{} Harris, W.E., Harris, G.L.H., Holland, S.T., McLaughlin, D.E. 2002, AJ, 124, 1435

\bibitem{} Harris, W.E., Whitmore, B.C., Karakla, D. et al. 2005, ApJ, in press 
(astro-ph/0508195)

\bibitem{} Ha\c{s}egan, M., Jord\'an, A., C\^ot\`e, P. et al. 2005, ApJ, 627, 203 

\bibitem{} Hempel, M., Kissler-Patig, M. 2004, A\&A, 428, 459

\bibitem{} Hilker, M., Infante, L., Vieira, G. et al. 1999,
A\&AS 134, 75

\bibitem{} Hilker, M. Richtler, T. 2000, A\&A, 362, 895

\bibitem{} Hunter, D.A., O'Connell, R.W., Gallagher, J.S., Smecker-Hane, T.A. 2000, AJ, 120, 2383

\bibitem{} Huxor, A.P., Tanvir, N.R., Irwin, M.J. et al. 2004, MNRAS, 260, 1007 

\bibitem{} Jordan, A., C\^ot\`e, P., Blakeslee, J.P. et al. 2005, ApJ, 624, 1002 
\bibitem{} Kissler-Patig, M., Richtler, T., Storm, J., della Valle, M. 1997, A\&A, 327, 503

\bibitem{} Kravtsov, A., Gnedin, O., 2005, ApJ, 623, 650

\bibitem{} Kundu, A., Whitmore, B.C. 2001a, AJ, 121, 2950 

\bibitem{} Kundu, A., Whitmore, B.C. 2001b, AJ, 122, 1251 

\bibitem{} Larsen, S.S. 2001, AJ, 122, 1782

\bibitem{} Larsen, S.S., Richtler, T.,  1999, A\&A, 122, 1782

\bibitem{} Larsen, S.S., Richtler, T.,  2000, A\&A, 122, 1782

\bibitem{} Larsen, S.S., Richtler, T.,  2004, A\&A, 427, 495

\bibitem{} Larsen, S.S., Brodie, J.P., Elmegreen, B.G., et al. 2001, ApJ, 567, 896

\bibitem{} Larsen, S.S., Efremov, Y.N., Elmegreen, B.G., et al. 2002, ApJ, 567, 896

\bibitem{} Maraston, C., Bastian, N., Saglia, R.P. et al. 2004, A\&A 416, 467

\bibitem{} Martini, P., Ho, L.C. 2004, ApJ, 610, 233

\bibitem {} Mieske, S., Hilker, M., Infante, L. 2002, A\&A,  383, 823 

\bibitem{} Mieske, S., Hilker, M., Infante , L. 2004, A\&A, 418, 445

\bibitem{} Mieske, S., Infante, L., Hilker, et al. 2005, A\&A, 430, L25

\bibitem{} Minniti, D., Kissler-Patig, M., Goudfrooij, P., Meylan, G. 1998, AJ, 115, 121

\bibitem{} Ostrov, P.G., Forte, J.C., Geisler, D. 1998, AJ, 116. 2854 

\bibitem{} Peng, E.W., Ford, H.C., Freeman, K. 2004, ApJ, 602, 705

\bibitem{} Peng, E.W., Jordan, A.,  C\^ot\`e, P. et al. 2005, ApJ, in press (astro-ph/0509654)

\bibitem{} Phillips, S., Drinkwater, M.J., Gregg, M.D, Jones, J.B. 2001, ApJ, 560, 201

\bibitem{} Pierce, M. Brodie, J.P., Forbes, D.A., et al. 2005,
MNRAS, 358, 419

\bibitem{} Piotto, G., Villanova, S., Bedin, L.R. et al. 2005, ApJ, 621, 777

\bibitem{} Puzia, T., Zepf, S.E, Kissler-Patig, M. et al. 2002, A\&A, 391, 453

\bibitem{} Richtler, Dirsch, B., Gebhardt, K., et al. 2004, AJ, 127, 2114  

\bibitem{} Richtler, T., Dirsch, B., Larsen, S.S., et al. 2005, A\&A, 439, 533

\bibitem{} Schweizer, F., Seitzer, P., Brodie, J.P. 2004,
AJ 128, 202

\bibitem{} Strader, J., Brodie, J.P., Forbes, D.A., 2004, AJ 127, 295

\bibitem{} Strader, J., Brodie, J.P., Spitler, L., Beasley, M.A. 2005a, submitted to AJ (astro-ph/0508001)

\bibitem{} Strader, J., Brodie, J.P, Cenarro, A.J., Beasley, M.A., Forbes, D.A.
2005b, AJ, 130, 1315 

\bibitem{} Tenorio-Tagle, G. Palous, J. Silich, S. et al. 2003,
A\&A, 411, 397 

\bibitem{} Tran, H.D., Sirianni, M., Ford, H.C., et al. 2003,
ApJ 585, 750

\bibitem{} van den Bergh, S., 1975, 
  ARAA, 13, 217

\bibitem{} Schweizer, F. 1987, in ''Nearly Normal Galaxies'', 8th Santa Cruz Summer Workshop,
New York, Springer, p.18

\bibitem{} Whitmore B.C., Schweizer F., Leitherer C., et al. 
1993, AJ, 106, 1354

\bibitem{} Whitmore, B.C., Schweizer, F. 1995,  AJ  109, 960

\bibitem{} Whitmore, B.C, Zhang, Q., Leitherer, C., Fall, S.M. 1999, AJ, 118, 1551

\bibitem{} Zepf, S.E., Ashman, K.M. 1993, MNRAS, 264, 611
\end{thebibliography}
\end{document}